\pdfoutput=1
\documentclass[twocolumn,showpacs,prl,superscriptaddress]{revtex4-1}
\usepackage{graphicx}
\usepackage{amsfonts}
\usepackage{amsmath} 
\usepackage{wasysym} 
\usepackage{amssymb} 
\usepackage{enumitem} 
\usepackage[english]{babel} 
\usepackage{color} 

\graphicspath{{figures/}}

\usepackage[]{hyperref}
\hypersetup{
  colorlinks   = true, 
  urlcolor     = blue, 
  linkcolor    = blue, 
  citecolor   = red 
}

\newcommand{\be}{\begin{equation}}
\newcommand{\ee}{\end{equation}}

\newcommand{\ben}{\begin{equation*}}
\newcommand{\een}{\end{equation*}}

\renewcommand{\d}[1]{\ensuremath{\operatorname{d}\!{#1}}}

\newcommand{\ane}{\ensuremath{\beta}}

\newcommand{\ps}{,}

\newcommand{\qex}{\ensuremath{q}}

\newcommand{\wte}{\ensuremath{\alpha}}

\newcommand{\wt}{{\tau}}

\newcommand{\lag}{t}

\newcommand{\dumlag}{{t'}}

\newcommand{\enst}{t}

\newcommand{\radius}{{r}}

\newcommand{\PD}{\ensuremath{{P_D}}}

\newcommand{\PDR}{\ensuremath{{P_{D,\radius}}}}

\newcommand{\PDT}{\ensuremath{{P_{D,\wt}}}}

\newcommand{\Ptau}{\ensuremath{{P_\wt}}}

\newcommand{\Pradius}{\ensuremath{{P_\radius}}}

\newcommand{\expect}[1]{\ensuremath{\mathrm{E}\left[{#1}\right]}}

\newcommand{\ensav}[1]{{\left\langle{#1}\right\rangle}}

\newcommand{\timeav}[2]{\overline{{#1}}_{{#2}}}

\newcommand{\xr}{{x}}
\newcommand{\xrsqp}{{x^2}}
\newcommand{\xrsq}[1]{{x^2(#1)}}

\newcommand{\tabr}[1]{Table~\ref{#1}}

\newcommand{\myheading}[1]{{\textit{#1}---}}

\begin{document}
\title{Nonergodic Subdiffusion from Brownian Motion in an Inhomogeneous Medium}
\newcommand{\icfoaf}{\affiliation{ICFO--Institut de Ci\`encies Fot\`oniques, Mediterranean
Technology Park, 08860 Castelldefels, Spain}}
\newcommand{\icreaaf}{\affiliation{ICREA-Instituci\'o Catalana de Recerca i Estudis Avan\c cats, Lluis Companys 23, 08010 Barcelona, Spain}}


\author{P. Massignan}
\icfoaf
\author{C. Manzo}
\icfoaf
\author{J. A. Torreno-Pina}
\icfoaf
\author{M. F. Garc\'{i}a-Parajo}
\icfoaf
\icreaaf
\author{M. Lewenstein}
\icfoaf
\icreaaf
\author{G. J. \surname{Lapeyre}, Jr.}
\icfoaf

\date{\today}

\begin{abstract}
  Non-ergodicity observed in single-particle tracking experiments is
  usually modeled by transient trapping rather than spatial
  disorder.  We introduce models of a particle diffusing in a medium
  consisting of regions with random sizes and random diffusivities.
  The particle is never trapped, but rather performs
  continuous Brownian motion with the local diffusion constant.  Under
  simple assumptions on the distribution of the sizes and
  diffusivities, we find that the mean squared displacement
  displays subdiffusion due to non-ergodicity for both
  annealed and quenched disorder. The model is formulated
  as a walk continuous in both time and space, similar to
  the L\'{e}vy walk.
\end{abstract}


\maketitle

Disordered systems exhibiting subdiffusion have been studied intensively
for decades~\cite{Havlin87,BG1990,Metzler2004,Klafter2011,Hoefling2013}. In these
systems the ensemble averaged mean squared displacement (EMSD) grows for
large times as
\begin{equation}
\ensav{x^2(\enst)} \sim \enst^\ane \quad \text{ with } 0<\ane<1,
 \label{defEMSD}
\end{equation}
whereas normal diffusion has $\ane=1$.
A broad class of systems show weak ergodicity breaking, that is,
the EMSD and the time averaged mean squared displacement (TMSD) differ.
The prototypical framework for describing non-ergodic subdiffusion is the 
heavy-tailed continuous-time random walk
(CTRW)~\cite{MontrollWeiss65,Scher1973,ScherMontroll75}, in which a particle
takes steps at random time intervals that are independently distributed with
density
\begin{equation}
 \psi(\wt) \sim \wt^{-\wte-1} \quad \quad 0<\wte<1.
 \label{ttdist}
\end{equation}
$\psi(\wt)$ has infinite mean,
which leads to a subdiffusive EMSD $\ane=\alpha$. 
Furthermore, the CTRW shows weak ergodicity breaking
because the particle experiences
trapping times on the order of the observation time $T$ no matter
how large $T$ is. 
The CTRW was introduced to describe charge carriers in amorphous
solids~\cite{ScherMontroll75}, and has found wide application since.
Recently, there has been a surge of work on the
CTRW~\cite{HeBMB2008,LubelskiSK2008,Meroz2010,Barkai2012a}, triggered by single
particle tracking experiments in biological
systems~\cite{Tolic04,Golding06,Jeon2011,Weigel2011,Kusumi2012} that
display signatures of non-ergodicity.

A different approach to subdiffusion is to assume a deterministic
diffusivity (\textit{i.e.} diffusion coefficient) that is
inhomogeneous in time~\cite{Saxton1993,Saxton1997}, or
space~\cite{Leyvraz1986,Hottovy12,Cherstvy2013a,Cherstvy2013b,Cherstvy2014}. But
in fact, the anomalous diffusion in these works is also non-ergodic.
Formulating models of inhomogeneous diffusivity is timely and
important, given that recently measured spatial maps in the cell
membrane often show patches of strongly varying
diffusivity~\cite{Serge2008,English2011,Kuehn2011,Cutler13,Giannone13,Masson2014}.
The presence of randomness in these experimental maps inspired us to
consider disordered media.  Thus, in this manuscript, we introduce
a class of models of ordinary diffusion with a diffusivity that varies
randomly but is constant on patches of random sizes. We call these
models random patch models or just \emph{patch models}.  These models
show non-ergodic subdiffusion, due to the diffusivity
effectively changing at random times with a heavy-tailed distribution
like that in~\eqref{ttdist}~\cite{[{Diffusion in a periodic potential with
    disorder that may show asymptotic non-ergodicity is considered in
  }]Khoury2011}. Note that ergodicity breaking
is usually ascribed to energetic disorder that immobilizes
the particle, e.g. via transient chemical
binding~\cite{ScherMontroll75,Yuste2007,Condamin2008}.  But, in the patch
models discussed here the particle constantly undergoes Brownian motion. The anomaly
is introduced not by transient immobilization, but rather by a
disordered medium.
This is a crucial distinction because, although
non-ergodicity and heterogeneity are often observed in the same system,
the toolbox for describing them is rather
spare~\cite{Hoefling2013}. Patch models address the pressing need
to enlarge this toolbox.


After introducing the models, we explain the origin of the
subdiffusivity \eqref{defEMSD}, and the dependence of the
exponent $\ane$ on the model parameters. Then we calculate $\ane$ for
a patch model using Fourier-Laplace techniques.  Next, we discuss the
conditions under which the linear behavior observed in the time-ensemble
averaged MSD (TEMSD) of the CTRW~\cite{HeBMB2008,LubelskiSK2008} may
occur in other models, and its appearance in patch models. Next we present our
numerical results. Finally, we address future work.

The disorder in these models is introduced via independent
and identically distributed pairs of random variables
$\{(D_j,\wt_j)\}$ or $\{(D_j,\radius_j)\}$. Here, $D_j$ is
a diffusivity, $\wt_j$ is a \emph{transit time}, and
$\radius_j$ is a length scale (radius).
For clarity, we concentrate on the one-dimensional case.

\myheading{Annealed transit time model (ATTM)}
In this model, the particle begins at $x=t=0$
and diffuses for a time $\wt_{1}$ with diffusivity $D_{1}$. 
Then, a new pair $(D_{2},\wt_{2})$ is sampled, and from time
$\wt_{1}$ to $\wt_{1}+\wt_{2}$ the particle diffuses with
diffusivity $D_{2}$. 
Diffusion then continues for the third pair and so on.
We assume that the pairs $\{(D_j,\wt_j)\}$ are
distributed with a probability density function (PDF)
$\PDT(D,\wt)=\PD(D)\Ptau(\wt|D)$, such that as $D\to 0$,
\begin{equation}
\PD(D) \sim D^{\sigma-1} \quad \quad \text{ with }
 \sigma>0,
\label{dmarginal}
\end{equation}
and that $\PD(D)$ decays rapidly for large $D$.
Furthermore, we require that the PDF
for $\wt$ given that we have sampled $D$, $\Ptau(\wt|D)$, has mean
\begin{equation}
\expect{\wt|D}=D^{-\gamma} \quad \quad \text{ with } -\infty<\gamma<\infty.
\label{wtconditional}
\end{equation}
\myheading{Annealed radius model (ARM)}
Here we take the radius $\radius_j$ to be random rather than $\wt_j$.
The particle begins at
the center of the first patch with $(D_1,\radius_1)$ and diffuses
until it hits the boundary of the patch, whereupon
a new patch with $(D_2,\radius_2)$ is sampled.
After hitting the boundary, the motion continues
at the \textit{center} of the new patch.
We take $\PDR(D,\radius)=\PD(D)\Pradius(\radius|D)$, where
$\Pradius(\radius|D)$ has mean $\expect{\radius|D}=D^{(1-\gamma)/2}$.
Since $\ensav{x^2(t)}\propto Dt$, this choice of the exponent ensures that
typical values of $r_j$ are the same as those of $\sqrt{D_j\wt_j}$.  As we
will see, the average behavior of the ARM and the ATTM is the same.  In the
annealed patch models, a new pair $(D_j,\radius_j)$ or $(D_j,\wt_j)$ is sampled
every time the particle hits a border.  An example of a system showing annealed disorder
is a protein subject to receptor-ligand interactions or conformational
changes that modulate the coupling with its
environment~\cite{Bakker2012,Rossier2012}. The result is a diffusivity that
is not associated with a position on the membrane, but rather fluctuates in
time.

\myheading{Quenched radius model (QRM)}
In this model, we have pairs $(D_j,\radius_j)$ with the
same PDF $\PDR(D,\radius)$ as in the ARM.
The difference is that the patches are fixed in space for
the duration of each trajectory. Thus, if the particle
crosses a border from patch $j$ with $(D_j,\radius_j)$ to patch $j+1$,
and later crosses back to patch $j$, it will find again the
same $(D_j,\radius_j)$. In fact, it may visit the same patch many times.
An example of a system with quenched disorder is
diffusion on liquid ordered/disordered phases of a lipid membrane~\cite{Sezgin2012}.
Depending on dimension and details of the model, 
the difference between quenched and annealed disorder may drastically
affect the dynamics. We found that this is indeed the case for the QRM compared to the
ATTM and ARM.

\begin{table}[t]
\begin{center}
   \begin{ruledtabular}
   \begin{tabular}{@{\extracolsep{10pt}} l c c c}  
     & (\textbf{0})  &  (\textbf{I}) & (\textbf{II}) \\
    & $\gamma<\sigma$ &   $\sigma<\gamma<\sigma+1$ & $\sigma +1 <\gamma $\\
   \hline
   Annealed      & 1 & $\sigma/\gamma$           & $1-1/\gamma$  \\
   Quenched 1d & 1 & $2\sigma/(\sigma+\gamma)$ & Unknown  \\
\end{tabular}
\end{ruledtabular}
\end{center}
\caption{\label{tab:anomexps}
EMSD exponents $\ane$ in \eqref{defEMSD} for the annealed (ATTM and ARM) and one-dimensional quenched (1d QRM) models, as a function of $\sigma$ and $\gamma$ defined in  \eqref{dmarginal} and
\eqref{wtconditional}. The exponent $\beta$ for the 1D QRM in region II is unknown at present.
}
\end{table}


\myheading{Anomalous exponents}
As we will see, all patch models exhibit a regime of normal diffusion
(\textbf{0}), and two anomalous regimes: (\textbf{I}) and (\textbf{II}).
The corresponding exponents are summarized in \tabr{tab:anomexps}, and will
be derived below.  Their origin however may be understood in simple terms
by considering the ATTM with the simplest PDF
satisfying \eqref{wtconditional}, $\Ptau(\wt|D) = \delta(\wt-D^{-\gamma})$,
that is $\wt=D^{-\gamma}$.
Using \eqref{dmarginal}, we find the PDF for the
transit time
\begin{equation}
\psi(\wt)\d{\wt} = \PD(D(\wt))\frac{\d{D}}{\d{\wt}}\d{\wt}
 \sim \wt^{-\frac{\sigma}{\gamma}-1} \d{\wt},
\label{psipatch}
\end{equation}
which has a heavy tail for $\sigma<\gamma$.
The density \eqref{psipatch} will play the role
of the waiting-time density \eqref{ttdist} with
$\alpha=\sigma/\gamma$.
In fact, if we observe the ATTM with
a stroboscope that illuminates the particle only at the
final position on each patch, we see exactly a CTRW
with waiting times $\wt_j=D_j^{-\gamma}$ and step lengths
with variance $\wt_j D_j = D_j^{(1-\gamma)/2}$.
Equivalently, we can generate $\wt=r^2/D$ from a random radius
$\radius=D^{(1-\gamma)/2}$ with
PDF $\Pradius(r) \sim r^{-\frac{2\sigma}{\gamma-1}-1}$, which has a diverging
variance when $\sigma+1<\gamma$. Similar arguments
for the ARM and QRM, as well as for the asymptotic forms of other
distributions for $\Ptau(\wt|D)$, and $\Pradius(\radius|D)$,
result in the same boundaries between regimes as in the ATTM.
These observations explain the regimes in \tabr{tab:anomexps}
showing that regime (\textbf{I}) corresponds to divergent $\expect{\wt}$ and finite
$\expect{\radius^2}$,
while in regime (\textbf{II}), both
$\expect{\wt}$ and $\expect{\radius^2}$ are divergent.
In this way, regime (\textbf{II}) is similar to the L\'{e}vy walk~\cite{Shlesinger1987,Klafter2011}.


\myheading{Fourier-Laplace transform solution} Here we compute
$\ensav{\xrsq{\enst}}$ in \eqref{defEMSD} for the ATTM using
techniques for analyzing CTRWs in which the waiting time
and the step length are not
independent~\cite{Klafter1987,Klafter2011}.
We again assume that the PDF for $\wt$ is concentrated on a point, \textit{i.e.}
$\wt=D^{-\gamma}$.
To describe partially completed motion on a patch,
we write the probability
density for a displacement $\xr$ at time $\wt$ on a patch with transit
time $\wt'$ such that $\wt\le \wt'$ \footnote{%
When writing probability densities and probabilities,
we do not distinguish between
arguments representing values of random variables
and other parameters. However, we do write the
former before the latter.}: 
\begin{equation}
  \psi(x,\wt'\ps \wt) = \phi(x|\wt'\ps \wt)\psi(\wt').
 \label{psixtt}
\end{equation}
We write the PDF for a
displacement $\xr$ at the end of a step, that is at time $\wt$, on a
patch with transit time $\wt$, as
\begin{equation}
  \phi(x|\wt)  \equiv \phi(x|\wt\ps \wt).
 \label{phixt}
\end{equation}
Likewise, $\psi(x,\wt)  \equiv \psi(x,\wt\ps \wt)$.
For the PDF of the displacement on a patch $\xr$ at time $\wt$,
when the only information we have on the transit time $\wt'$
is $\wt<\wt'$, we write
\begin{equation}
\Psi(x\ps \wt) = \int_\wt^\infty \psi(x,\wt'\ps \wt) \d{\wt'}.
 \label{Psidef}
\end{equation}
$\Psi(x\ps\wt)$ describes the displacement of the particle on the final,
uncompleted, patch.  Note that if $\phi(x|\wt'\ps\wt)$ is independent of
$\wt'$, we have $\Psi(x\ps\wt) = \phi(x|\wt'\ps\wt) \Psi(\wt)$, where
the survival probability $\Psi(\wt)=\int_\wt^\infty \psi(\wt')\d{\wt'}$ is the probability that
a step is not completed by time $\wt$. An example is the L\'{e}vy
walk~\cite{Klafter1987,Klafter2011}, in which the walker undergoes
rectilinear motion on each step; that is, $\psi(x,\wt'\ps\wt)=
\delta(|x|-c\wt) \psi(\wt')$, where the speed $c$ is independent of
$\wt'$. In our case, however, $D$ is not independent of $\wt'$
and this simplification cannot be made.

We denote by $P(x\ps\enst)$ the PDF for the particle to be at $x$ at time $\enst$,
with the initial condition $P(x\ps\enst=0)=\delta(x)$, and by
$\eta(x\ps t)$ the PDF of
the particle's position at time $t$ just after having completed a step.
Then
$\eta(x\ps t) = \delta(x)\delta(t) + \int_{-\infty}^\infty \d{x'} \int_0^t \d{t'}\eta(x'\ps t') \psi(x-x'\ps t-t')$
and
$P(x\ps t) = \int_{-\infty}^\infty \d{x'} \int_0^t \d{t'} 
 \eta(x'\ps t') \Psi(x-x'\ps t-t').$
The Fourier-Laplace representation of $P(x\ps\enst)$ is~\cite{Klafter2011}
\begin{equation}
 P(k\ps s) = \frac{\Psi(k\ps s)}{1-\psi(k,s)},
 \label{Pks}
\end{equation}
where $\Psi(k\ps s)$ is the transform of $\Psi(x\ps \wt)$, and
likewise with $\psi(k,s)$ and $\psi(x,\wt)$. 
We compute only the second moment of $P(x\ps \enst)$,
which reads in Laplace space
\begin{equation}
  \ensav{x^2(s)} = -P''(k\ps s)|_{k=0},
 \label{xsqs}
\end{equation}
where prime means differentiation with respect to $k$.
It is easy to see that 
$\psi(k=0,s)=\psi(s)$ and $\Psi(k=0\ps s)=\Psi(s)$.
Moreover, the first moments $\psi'(k=0,s)$ and  $\Psi'(k=0\ps s)$ vanish because the diffusion is unbiased.
Using \eqref{Pks}, \eqref{xsqs}, and $\Psi(s)=[1-\psi(s)]/s$~\cite{Klafter2011}, we obtain
for generic $\psi(x,t)$
\begin{equation}
 \ensav{x^2(s)} = \frac{-\psi''(k,s)|_{k=0}}{s [1-\psi(s)]}
 +  \frac{-\Psi''(k\ps s)|_{k=0}}{1-\psi(s)}.
 \label{genmsd}
\end{equation}
If the particle does not move during
the transit times, but only jumps at the end of each one,
as in the CTRW, then
the second term in \eqref{genmsd} vanishes.
Now we assume a heavy tailed transit-time density
\eqref{ttdist}, which has Laplace transform
$\psi(s)\sim 1-b s^\wte$ for small $s$~\cite{Klafter2011}, so that
for small~$s$ (corresponding to large $t$) \eqref{genmsd} becomes
\begin{equation}
 \ensav{x^2(s)} \sim \frac{-\psi''(k,s)|_{k=0}}{s^{\wte+1}}
 +  \frac{-\Psi''(k\ps s)|_{k=0}}{s^\wte}.
 \label{agenmsd}
\end{equation}
We now specialize to the ATTM, whose displacements obey the Brownian
propagator
\begin{equation}
\phi(x|\wt'\ps\wt) = \frac{1}{\sqrt{2\pi D(\wt') \wt}}\exp\left(\frac{-\xrsqp}{2 D(\wt') \wt} \right),
 \label{phibrownian}
\end{equation}
with $D(\wt)=\wt^{-1/\gamma}$.
We first consider the PDF \eqref{phixt} of $x$ at the end of a step.
For clarity, we write
$f(\wt)$ for $D(\wt)\wt$, and suppose $f(\wt)\sim \wt^\qex$.
Then, using \eqref{phixt} and \eqref{phibrownian}
the Fourier transform of $\phi(x|\wt)$ is
$\phi(k|\wt) =   \exp\left(-k^2 f(\wt)/2\right)$,
so that $\phi''(k|\wt)|_{k=0}= -f(\wt)\sim -\wt^\qex$.
Combining this  with \eqref{ttdist}, \eqref{psixtt} and \eqref{phixt},
we see that $\psi''(k,\wt)|_{k=0}\sim \wt^{\qex-\wte-1}$.
If $0<\wte<1$ and $\qex>\wte$, then a Tauberian theorem~\cite{feller-vol-2,Klafter2011} 
gives $\psi''(k,s)|_{k=0}\sim s^{\wte-\qex}$. Thus, the first term
in \eqref{agenmsd} becomes $s^{-\qex-1}$.
Using \eqref{ttdist}, \eqref{psixtt}, \eqref{Psidef},
and \eqref{phibrownian},  we find 
\begin{equation*}
-\Psi''(k\ps\wt)|_{k=0} = \wt \int_\wt^\infty \!\!\!\! D(\wt')\psi(\wt') \d{\wt'}
  \sim \wt \int_\wt^\infty \!\!\!\! {\wt'}^{\qex-\wte-2} \d{\wt'}.
\end{equation*}
Performing the integral, applying the Tauberian theorem, and
inserting the result in \eqref{agenmsd}, we find that the
second term scales with the same exponent as the first.
Thus, accounting for the continuous motion
does not affect the EMSD, which remains the same as in the CTRW.
The inverse Laplace transform of \eqref{agenmsd}
gives us
\begin{equation}
\ensav{x^2(\enst)} \sim \enst^\qex  \quad \text{ for } \qex > \wte.
\label{asympone}
\end{equation}
Now we consider the case $\qex<\wte$.
$\psi''(k,s)|_{k=0}$ no longer satisfies
the hypothesis of the Tauberian theorem.
But its integral does, which
leads to
$\psi''(k,s)|_{k=0}\sim c-bs^{\wte-\qex}$. Thus, the first term
in \eqref{agenmsd} is
$\ensav{x^2(s)} \sim (c-b s^{\wte-\qex})/s^{\wte+1}$,
or for small $s$, $\ensav{x^2(s)} \sim {s^{-\wte-1}}$.
A similar calculation again shows that the second term has the same
exponents. 
The inverse Laplace transform gives
\begin{equation}
\ensav{x^2(\enst)} \sim \enst^\wte  \quad \text{ for } \qex < \wte.
\label{asymptwo}
\end{equation} 
We return now to the ATTM, recalling that $f(\wt)=D(\wt)\wt\sim
\wt^{1-1/\gamma}$ so that $\qex=1-1/\gamma$. Using \eqref{psipatch} for \eqref{ttdist}
we have $\wte=\sigma/\gamma$. Thus \eqref{asympone}
becomes $\ensav{x^2(\enst)} \sim \enst^{1-1/\gamma}$ for $\gamma>\sigma+1$,
and \eqref{asymptwo} becomes $\ensav{x^2(\enst)} \sim
\enst^{\sigma/\gamma}$ for $0<\sigma<\gamma$.
Note that these two conditions on $\sigma$ and $\gamma$ are exactly those
defining the anomalous regimes in the discussion following
\eqref{psipatch}. 
The value of $\ane$ for the QRM in regime (\textbf{I}) is explained
by comparison with the quenched version of the CTRW, in which trapping
times are assigned to sites on a
lattice.  In one dimension, the exponent
of the EMSD \eqref{defEMSD} for the quenched CTRW
with the waiting time PDF \eqref{ttdist}
is $\ane=2\alpha/(1+\alpha)$~\cite{Machta85,BG1990,Bouchaud1992}. Substituting
$\alpha=\sigma/\gamma$, we find $\beta=2\sigma/(\sigma+\gamma)$.


\myheading{Time-ensemble averaged MSD}
It is becoming clear that the TEMSD
is important both theoretically and
as an experimental tool for elucidating the source
of subdiffusion\cite{HeBMB2008,LubelskiSK2008,Meroz2010,Weigel2011}. 
The TEMSD is given by
\begin{equation}
  \timeav{\ensav{\xrsq{\lag}}}{T} 
    = \frac{1}{T-\lag}\int_0^{T-\lag} \ensav{\left[\xr(\lag+\dumlag)
      -\xr(\dumlag)\right]^2} \d{{\dumlag}},
  \label{defTEMSD}
\end{equation}
where $\lag$ is the time lag, $T$ the observation time,
and the overbar denotes the time average.
Suppose $x(\enst)$ is a process with mean zero and
that the EMSD and TMSD exist.
If $x(\enst)$ has stationary increments in the wide sense,
then the integrand in \eqref{defTEMSD} is independent of $\lag'$,
and we have that $\timeav{\ensav{\xrsq{\lag}}}{T}=\ensav{\xrsq{\lag}}$~\cite{Durrett96}.
Let us now consider $x(\lag)$ without the restriction to stationary increments.
Expanding the integrand in \eqref{defTEMSD} and rearranging the limits on
the integrals we find
\begin{multline}
\timeav{\ensav{x^2(\lag)}}{T}
 = \frac{1}{T-\enst}\int_{T-\enst}^{T} \ensav{x^2(\lag')}\d{\lag'} \\
 - \frac{1}{T-\lag}\int_{0}^{\lag} \ensav{x^2(\lag')}\d{\lag'} 
  - \frac{2}{T-\lag}\int_{0}^{T-\lag} g(\lag,\lag')\d{\lag'},
\end{multline}
where $g(\lag,\lag')=\ensav{[x(\lag+\lag')-x(\lag')]x(\lag')}$ is the
correlation between the increments $x(\lag+\lag')-x(\lag')$ and
$x(\lag')-x(0)$. Now we assume that $g(\lag,\lag')=0$, that is,
$x(\lag)$ has uncorrelated increments.  Then the third term vanishes.
We furthermore
assume that $t\ll T$ and that $\ensav{x^2(\lag)}$ continues to
increase with increasing $\lag$. Then the second term vanishes more
rapidly than the first with increasing $T$.
Thus, the dominant contribution comes from the
time interval  $[T-\lag,T]$.
Finally, if the EMSD is subdiffusive as in \eqref{defEMSD}
then the first term becomes $T^{\ane-1}\lag$.
Thus, if
\renewcommand{\labelenumi}{(\textit{\roman{enumi}})}
\begin{enumerate}
 \item $x(\enst)$ has uncorrelated increments
\end{enumerate}
and
\begin{enumerate}[resume]
 \item $\ensav{x^2(\enst)} \sim \enst^\ane$ \quad  with $\ane\ne 1$,
\end{enumerate}
then $x(\enst)$ has non-stationary increments, it shows weak-ergodicity breaking, and its MSD satisfies
\begin{equation}
\timeav{\ensav{x^2(\enst)}}{T}\sim T^{\ane-1}\enst.
\label{indTEMSD}
\end{equation}
Brownian motion satisfies (\emph{i}), but not (\emph{ii}).
Both fractional Brownian motion~\cite{Mandelbrot1968} with $\ane<1$ and the
random walk on a fractal~\cite{Havlin87} satisfy (\emph{ii}), but not
(\emph{i}). The CTRW satisfies both (\emph{i})~\cite{Barkai2007} and
(\emph{ii}). The CTRW on a fractal satisfies (\emph{ii}), but not
(\emph{i}). It also shows non-ergodicity, but
$\timeav{\ensav{x^2(\enst)}}{T}\nsim
T^{\ane-1}\enst$~\cite{Meroz2010}.  The CTRW has been shown to
follow~\eqref{indTEMSD}~\cite{HeBMB2008,LubelskiSK2008}.  Furthermore,
the statistics of the time average $\timeav{x^2(\enst)}{T}$ for the
CTRW, which does not converge to a constant random variable, have been
studied in Ref.~\cite{HeBMB2008}.  We do not present a proof that
patch models satisfy (\emph{i}), but, in fact, our numerical results
show they follow~\eqref{indTEMSD}.

\begin{figure}[t!]
   \includegraphics[width= \columnwidth]{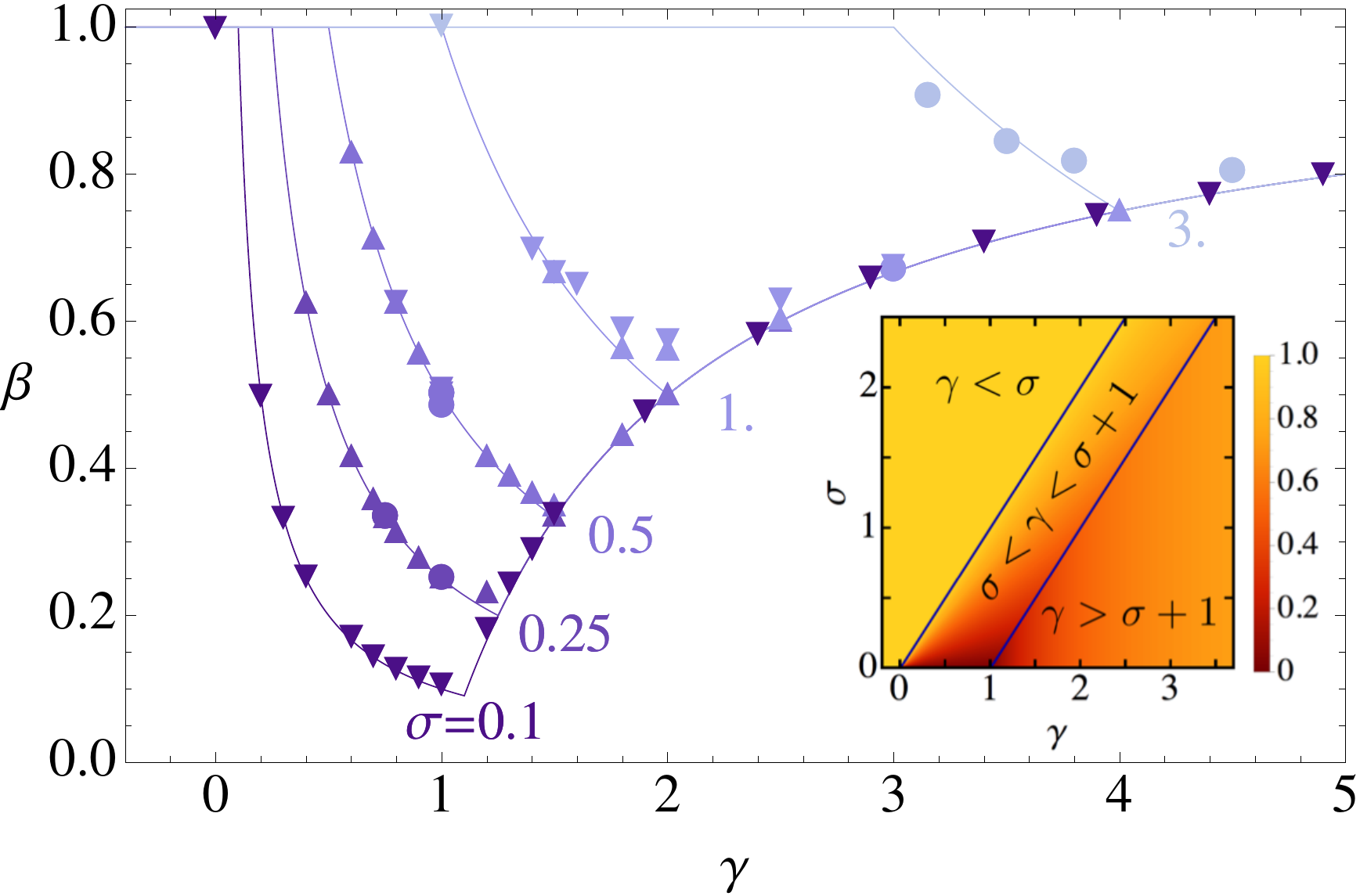}
   \caption{ Exponent $\ane$ in \eqref{defEMSD} for annealed models. Lines are
     analytic results as in \tabr{tab:anomexps}, for different values of $\sigma$ as
    indicated in the figure. Symbols are numerical simulations.
    Lines and symbols vary from dark to light with increasing $\sigma$.
    Exponents are extracted respectively from:
    EMSD of the ATTM  ($\blacktriangledown$),
    EMSD of the ARM ($\blacktriangle$)
    and TEMSD of the ARM ($\CIRCLE$).
     The inset shows a density plot of $\ane$ vs.  both $\gamma$ and $\sigma$.
  }
\label{fig:temsdAnnealed}
\end{figure}



\begin{figure}[t!]
   \includegraphics[width= \columnwidth]{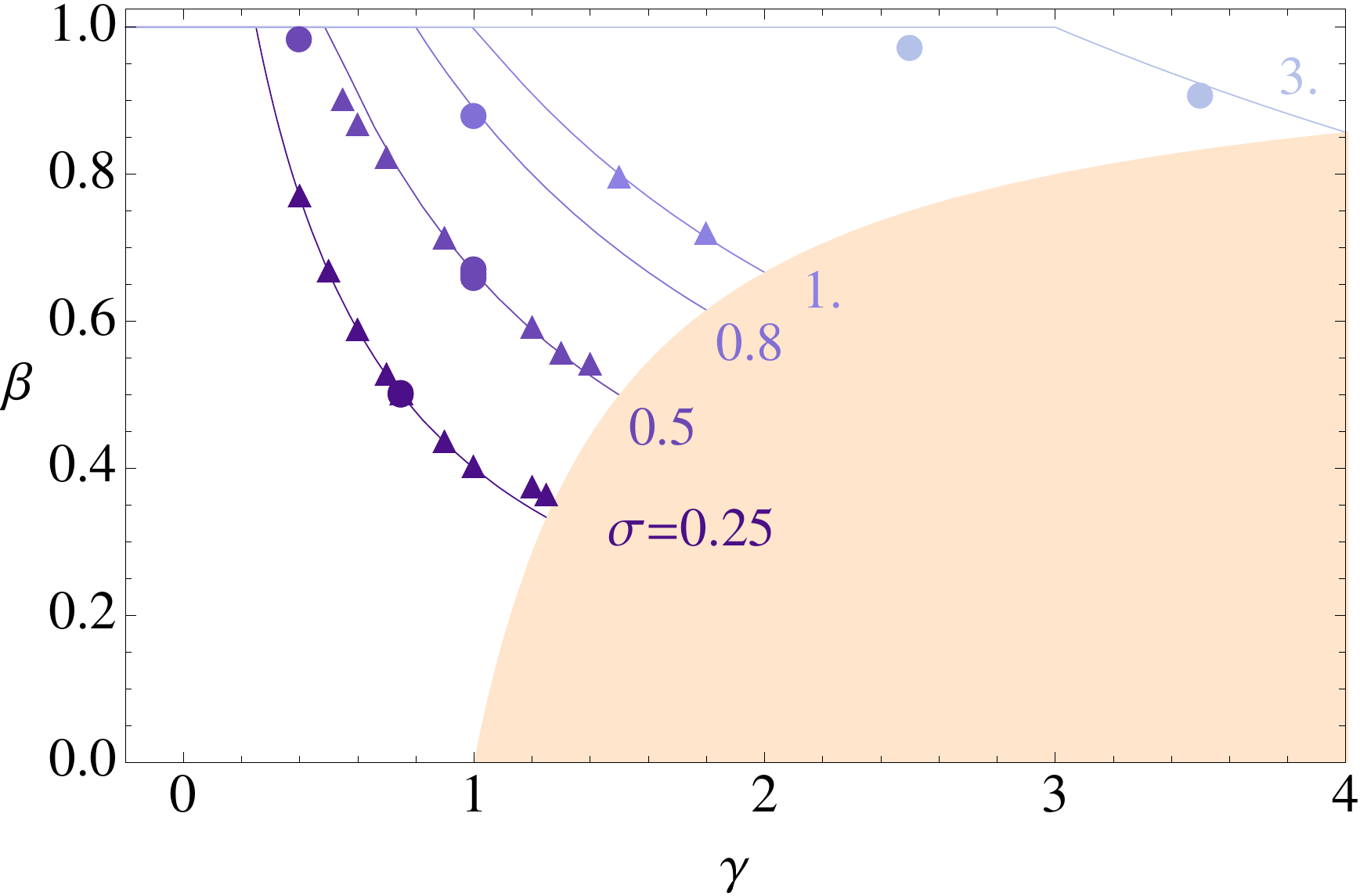}
   \caption{ Exponent $\ane$ in \eqref{defEMSD} for the 1d quenched radius model (1d QRM).
     Lines as in Fig.~\ref{fig:temsdAnnealed}. Symbols are
     exponents extracted from numerical simulations of the 
     EMSD ($\blacktriangle$) and TEMSD
     ($\CIRCLE$).
     Lines and symbols vary from dark to light with increasing $\sigma$.
     Shading indicates region ({\bf II}), where the
     exponent is at present unknown.}
\label{fig:temsdQuenched}
\end{figure}

\myheading{Simulations} The results of our extensive computer simulations of
all models are shown in Figs.~\ref{fig:temsdAnnealed}
and~\ref{fig:temsdQuenched}.  We used the gamma distribution for $\PD(D)$
in \eqref{dmarginal}, and (normal and stretched) exponential, log-normal,
and single-point distributions for $\PDT(D,\wt)$ and $\PDR(D,\radius)$.
The exponent $\ane$ was determined for the EMSD by a linear fit
of $\log[\ensav{x^2(\enst)}]$ \textit{vs} $\log(t)$.
To analyze the TEMSD, we first determined the diffusivities
by a linear fit of the TEMSD \textit{vs}.\ the lag at given $T$. We then
did a linear fit to a log-log plot of the resulting diffusivities \textit{vs.}
$T$ to get $\ane-1$ in \eqref{indTEMSD}. The exponents $\ane$ obtained
from the EMSD and TEMSD are in excellent agreement with
\tabr{tab:anomexps}.
The QRM in regime (\textbf{II}) clearly shows
subdiffusion. But at present we have no explanation for $\ane$ in this regime.

To understand why, in the ATTM, we position the particle at the center of a
new patch upon hitting a border, recall that a $1$d
Brownian path crosses a point infinitely many times before leaving
any neighborhood~\cite{Durrett96}. Now, assume annealed
disorder and that the particle enters a new patch at its boundary, as in
the QRM.  Because a new patch is sampled each time the border is crossed,
the particle samples an infinite number of patches during the crossing.  In
this case, our simulations of the EMSD did not converge with decreasing
step length.  But the EMSD does converge for the QRM, which visits the
same two patches an infinite number of times on crossing a border.

\myheading{Outlook and Applications} Many questions remain to be
addressed. For instance, what is the behavior at the boundaries of the
parameter regimes, that is for $\gamma=\sigma$ and $\gamma=\sigma+1$,
as well as in regime (\textbf{II}) for the QRM?
Regarding dimensions $d>1$: The ATTM and ARM are the same for all $d$, and
the EMSD for the quenched CTRW for $d>1$ has the same exponent $\ane$ as
the (annealed) CTRW, with logarithmic corrections for
$d=2$~\cite{BG1990,benarous2007}. But before analyzing the QRM in $d>1$, a geometry
of patches consistent with $\PDR(D,\radius)$ must be found.

Patch models provide an alternative for describing non-ergodic
diffusion in biological systems, one that is due to inhomogeneous
diffusivity rather than transient trapping. But there are many similarities
in the long-time behavior of the CTRW and the patch models.  Thus, the main open
problem is finding methods to distinguish them, and regime
(\textbf{I}) from (\textbf{II}). Promising leads in this
direction are studying a first-passage quantity such as the survival time density, or comparing the exponents $\sigma,\,\gamma$ and $\beta$ appearing in our models with those extracted from spatial maps of diffusivity and time-resolved trajectories, or performing a detailed analysis of the models in terms of trajectories with long but finite (i.e., not asymptotically long) duration.

\begin{acknowledgements}
\myheading{Acknowledgments}
   We acknowledge insightful discussions with Jan Wehr and Ignacio
   Izeddin. This work was supported by ERC AdG Osyris, Spanish Ministry
   of Science and Innovation (Grants No. FIS2008-00784 and
   MAT2011-22887), Generalitat de Catalunya (Grant No. 2009 SGR 597),
   Fundaci\'o Cellex, the European Commission (FP7-ICT-2011-7,
   Grant No. 288263), and the HFSP (Grant No. RGP0027/2012)
\end{acknowledgements}

\bibliography{Ddist}

\begin{thebibliography}{46}%
\makeatletter
\providecommand \@ifxundefined [1]{%
 \@ifx{#1\undefined}
}%
\providecommand \@ifnum [1]{%
 \ifnum #1\expandafter \@firstoftwo
 \else \expandafter \@secondoftwo
 \fi
}%
\providecommand \@ifx [1]{%
 \ifx #1\expandafter \@firstoftwo
 \else \expandafter \@secondoftwo
 \fi
}%
\providecommand \natexlab [1]{#1}%
\providecommand \enquote  [1]{``#1''}%
\providecommand \bibnamefont  [1]{#1}%
\providecommand \bibfnamefont [1]{#1}%
\providecommand \citenamefont [1]{#1}%
\providecommand \href@noop [0]{\@secondoftwo}%
\providecommand \href [0]{\begingroup \@sanitize@url \@href}%
\providecommand \@href[1]{\@@startlink{#1}\@@href}%
\providecommand \@@href[1]{\endgroup#1\@@endlink}%
\providecommand \@sanitize@url [0]{\catcode `\\12\catcode `\$12\catcode
  `\&12\catcode `\#12\catcode `\^12\catcode `\_12\catcode `\%12\relax}%
\providecommand \@@startlink[1]{}%
\providecommand \@@endlink[0]{}%
\providecommand \url  [0]{\begingroup\@sanitize@url \@url }%
\providecommand \@url [1]{\endgroup\@href {#1}{\urlprefix }}%
\providecommand \urlprefix  [0]{URL }%
\providecommand \Eprint [0]{\href }%
\providecommand \doibase [0]{http://dx.doi.org/}%
\providecommand \selectlanguage [0]{\@gobble}%
\providecommand \bibinfo  [0]{\@secondoftwo}%
\providecommand \bibfield  [0]{\@secondoftwo}%
\providecommand \translation [1]{[#1]}%
\providecommand \BibitemOpen [0]{}%
\providecommand \bibitemStop [0]{}%
\providecommand \bibitemNoStop [0]{.\EOS\space}%
\providecommand \EOS [0]{\spacefactor3000\relax}%
\providecommand \BibitemShut  [1]{\csname bibitem#1\endcsname}%
\let\auto@bib@innerbib\@empty
\bibitem [{\citenamefont {Havlin}\ and\ \citenamefont
  {Ben-Avraham}(1987)}]{Havlin87}%
  \BibitemOpen
  \bibfield  {author} {\bibinfo {author} {\bibfnamefont {S.}~\bibnamefont
  {Havlin}}\ and\ \bibinfo {author} {\bibfnamefont {D.}~\bibnamefont
  {Ben-Avraham}},\ }\href {\doibase 10.1080/00018738700101072} {\bibfield
  {journal} {\bibinfo  {journal} {Adv. Phys.}\ }\textbf {\bibinfo {volume}
  {36}},\ \bibinfo {pages} {695} (\bibinfo {year} {1987})}\BibitemShut
  {NoStop}%
\bibitem [{\citenamefont {{Bouchaud}}\ and\ \citenamefont
  {{Georges}}(1990)}]{BG1990}%
  \BibitemOpen
  \bibfield  {author} {\bibinfo {author} {\bibfnamefont {J.-P.}\ \bibnamefont
  {{Bouchaud}}}\ and\ \bibinfo {author} {\bibfnamefont {A.}~\bibnamefont
  {{Georges}}},\ }\href {\doibase 10.1016/0370-1573(90)90099-N} {\bibfield
  {journal} {\bibinfo  {journal} {Phys. Rep.}\ }\textbf {\bibinfo {volume}
  {195}},\ \bibinfo {pages} {127} (\bibinfo {year} {1990})}\BibitemShut
  {NoStop}%
\bibitem [{\citenamefont {Metzler}\ and\ \citenamefont
  {Klafter}(2004)}]{Metzler2004}%
  \BibitemOpen
  \bibfield  {author} {\bibinfo {author} {\bibfnamefont {R.}~\bibnamefont
  {Metzler}}\ and\ \bibinfo {author} {\bibfnamefont {J.}~\bibnamefont
  {Klafter}},\ }\href {http://stacks.iop.org/0305-4470/37/i=31/a=R01}
  {\bibfield  {journal} {\bibinfo  {journal} {J. Phys. A}\ }\textbf {\bibinfo
  {volume} {37}},\ \bibinfo {pages} {R161} (\bibinfo {year}
  {2004})}\BibitemShut {NoStop}%
\bibitem [{\citenamefont {Klafter}\ and\ \citenamefont
  {Sokolov}(2011)}]{Klafter2011}%
  \BibitemOpen
  \bibfield  {author} {\bibinfo {author} {\bibfnamefont {J.}~\bibnamefont
  {Klafter}}\ and\ \bibinfo {author} {\bibfnamefont {I.~M.}\ \bibnamefont
  {Sokolov}},\ }\href@noop {} {\emph {\bibinfo {title} {First Steps in Random
  Walks}}}\ (\bibinfo  {publisher} {Oxford University Press},\ \bibinfo
  {address} {Oxford},\ \bibinfo {year} {2011})\BibitemShut {NoStop}%
\bibitem [{\citenamefont {H\"{o}fling}\ and\ \citenamefont
  {Franosch}(2013)}]{Hoefling2013}%
  \BibitemOpen
  \bibfield  {author} {\bibinfo {author} {\bibfnamefont {F.}~\bibnamefont
  {H\"{o}fling}}\ and\ \bibinfo {author} {\bibfnamefont {T.}~\bibnamefont
  {Franosch}},\ }\href {\doibase 10.1088/0034-4885/76/4/046602} {\bibfield
  {journal} {\bibinfo  {journal} {Rep. Prog. Phys.}\ }\textbf {\bibinfo
  {volume} {76}},\ \bibinfo {pages} {046602} (\bibinfo {year}
  {2013})}\BibitemShut {NoStop}%
\bibitem [{\citenamefont {Montroll}\ and\ \citenamefont
  {Weiss}(1965)}]{MontrollWeiss65}%
  \BibitemOpen
  \bibfield  {author} {\bibinfo {author} {\bibfnamefont {E.~W.}\ \bibnamefont
  {Montroll}}\ and\ \bibinfo {author} {\bibfnamefont {G.~H.}\ \bibnamefont
  {Weiss}},\ }\href {\doibase http://dx.doi.org/10.1063/1.1704269} {\bibfield
  {journal} {\bibinfo  {journal} {J. Math Phys.}\ }\textbf {\bibinfo {volume}
  {6}},\ \bibinfo {pages} {167} (\bibinfo {year} {1965})}\BibitemShut {NoStop}%
\bibitem [{\citenamefont {Scher}\ and\ \citenamefont {Lax}(1973)}]{Scher1973}%
  \BibitemOpen
  \bibfield  {author} {\bibinfo {author} {\bibfnamefont {H.}~\bibnamefont
  {Scher}}\ and\ \bibinfo {author} {\bibfnamefont {M.}~\bibnamefont {Lax}},\
  }\href {\doibase 10.1103/PhysRevB.7.4491} {\bibfield  {journal} {\bibinfo
  {journal} {Phys. Rev. B}\ }\textbf {\bibinfo {volume} {7}},\ \bibinfo {pages}
  {4491} (\bibinfo {year} {1973})}\BibitemShut {NoStop}%
\bibitem [{\citenamefont {Scher}\ and\ \citenamefont
  {Montroll}(1975)}]{ScherMontroll75}%
  \BibitemOpen
  \bibfield  {author} {\bibinfo {author} {\bibfnamefont {H.}~\bibnamefont
  {Scher}}\ and\ \bibinfo {author} {\bibfnamefont {E.~W.}\ \bibnamefont
  {Montroll}},\ }\href {\doibase 10.1103/PhysRevB.12.2455} {\bibfield
  {journal} {\bibinfo  {journal} {Phys. Rev. B}\ }\textbf {\bibinfo {volume}
  {12}},\ \bibinfo {pages} {2455} (\bibinfo {year} {1975})}\BibitemShut
  {NoStop}%
\bibitem [{\citenamefont {He}\ \emph {et~al.}(2008)\citenamefont {He},
  \citenamefont {Burov}, \citenamefont {Metzler},\ and\ \citenamefont
  {Barkai}}]{HeBMB2008}%
  \BibitemOpen
  \bibfield  {author} {\bibinfo {author} {\bibfnamefont {Y.}~\bibnamefont
  {He}}, \bibinfo {author} {\bibfnamefont {S.}~\bibnamefont {Burov}}, \bibinfo
  {author} {\bibfnamefont {R.}~\bibnamefont {Metzler}}, \ and\ \bibinfo
  {author} {\bibfnamefont {E.}~\bibnamefont {Barkai}},\ }\href {\doibase
  10.1103/PhysRevLett.101.058101} {\bibfield  {journal} {\bibinfo  {journal}
  {Phys. Rev. Lett.}\ }\textbf {\bibinfo {volume} {101}},\ \bibinfo {pages}
  {058101} (\bibinfo {year} {2008})}\BibitemShut {NoStop}%
\bibitem [{\citenamefont {Lubelski}\ \emph {et~al.}(2008)\citenamefont
  {Lubelski}, \citenamefont {Sokolov},\ and\ \citenamefont
  {Klafter}}]{LubelskiSK2008}%
  \BibitemOpen
  \bibfield  {author} {\bibinfo {author} {\bibfnamefont {A.}~\bibnamefont
  {Lubelski}}, \bibinfo {author} {\bibfnamefont {I.~M.}\ \bibnamefont
  {Sokolov}}, \ and\ \bibinfo {author} {\bibfnamefont {J.}~\bibnamefont
  {Klafter}},\ }\href {\doibase 10.1103/PhysRevLett.100.250602} {\bibfield
  {journal} {\bibinfo  {journal} {Phys. Rev. Lett.}\ }\textbf {\bibinfo
  {volume} {100}},\ \bibinfo {pages} {250602} (\bibinfo {year}
  {2008})}\BibitemShut {NoStop}%
\bibitem [{\citenamefont {Meroz}\ \emph {et~al.}(2010)\citenamefont {Meroz},
  \citenamefont {Sokolov},\ and\ \citenamefont {Klafter}}]{Meroz2010}%
  \BibitemOpen
  \bibfield  {author} {\bibinfo {author} {\bibfnamefont {Y.}~\bibnamefont
  {Meroz}}, \bibinfo {author} {\bibfnamefont {I.~M.}\ \bibnamefont {Sokolov}},
  \ and\ \bibinfo {author} {\bibfnamefont {J.}~\bibnamefont {Klafter}},\ }\href
  {\doibase 10.1103/PhysRevE.81.010101} {\bibfield  {journal} {\bibinfo
  {journal} {Phys. Rev. E}\ }\textbf {\bibinfo {volume} {81}},\ \bibinfo
  {pages} {010101} (\bibinfo {year} {2010})}\BibitemShut {NoStop}%
\bibitem [{\citenamefont {Barkai}\ \emph {et~al.}(2012)\citenamefont {Barkai},
  \citenamefont {Garini},\ and\ \citenamefont {Metzler}}]{Barkai2012a}%
  \BibitemOpen
  \bibfield  {author} {\bibinfo {author} {\bibfnamefont {E.}~\bibnamefont
  {Barkai}}, \bibinfo {author} {\bibfnamefont {Y.}~\bibnamefont {Garini}}, \
  and\ \bibinfo {author} {\bibfnamefont {R.}~\bibnamefont {Metzler}},\ }\href
  {\doibase 10.1063/PT.3.1677} {\bibfield  {journal} {\bibinfo  {journal}
  {Phys. Today}\ }\textbf {\bibinfo {volume} {65}},\ \bibinfo {pages} {29}
  (\bibinfo {year} {2012})}\BibitemShut {NoStop}%
\bibitem [{\citenamefont {Toli\'{c}-N\o{}rrelykke}\ \emph
  {et~al.}(2004)\citenamefont {Toli\'{c}-N\o{}rrelykke}, \citenamefont
  {Munteanu}, \citenamefont {Thon}, \citenamefont {Oddershede},\ and\
  \citenamefont {Berg-S\o{}rensen}}]{Tolic04}%
  \BibitemOpen
  \bibfield  {author} {\bibinfo {author} {\bibfnamefont {I.~M.}\ \bibnamefont
  {Toli\'{c}-N\o{}rrelykke}}, \bibinfo {author} {\bibfnamefont {E.-L.}\
  \bibnamefont {Munteanu}}, \bibinfo {author} {\bibfnamefont {G.}~\bibnamefont
  {Thon}}, \bibinfo {author} {\bibfnamefont {L.}~\bibnamefont {Oddershede}}, \
  and\ \bibinfo {author} {\bibfnamefont {K.}~\bibnamefont {Berg-S\o{}rensen}},\
  }\href {\doibase 10.1103/PhysRevLett.93.078102} {\bibfield  {journal}
  {\bibinfo  {journal} {Phys. Rev. Lett.}\ }\textbf {\bibinfo {volume} {93}},\
  \bibinfo {pages} {078102} (\bibinfo {year} {2004})}\BibitemShut {NoStop}%
\bibitem [{\citenamefont {Golding}\ and\ \citenamefont
  {Cox}(2006)}]{Golding06}%
  \BibitemOpen
  \bibfield  {author} {\bibinfo {author} {\bibfnamefont {I.}~\bibnamefont
  {Golding}}\ and\ \bibinfo {author} {\bibfnamefont {E.~C.}\ \bibnamefont
  {Cox}},\ }\href {\doibase 10.1103/PhysRevLett.96.098102} {\bibfield
  {journal} {\bibinfo  {journal} {Phys. Rev. Lett.}\ }\textbf {\bibinfo
  {volume} {96}},\ \bibinfo {pages} {098102} (\bibinfo {year}
  {2006})}\BibitemShut {NoStop}%
\bibitem [{\citenamefont {Jeon}\ \emph {et~al.}(2011)\citenamefont {Jeon},
  \citenamefont {Tejedor}, \citenamefont {Burov}, \citenamefont {Barkai},
  \citenamefont {Selhuber-Unkel}, \citenamefont {Berg-S\o{}rensen},
  \citenamefont {Oddershede},\ and\ \citenamefont {Metzler}}]{Jeon2011}%
  \BibitemOpen
  \bibfield  {author} {\bibinfo {author} {\bibfnamefont {J.-H.}\ \bibnamefont
  {Jeon}}, \bibinfo {author} {\bibfnamefont {V.}~\bibnamefont {Tejedor}},
  \bibinfo {author} {\bibfnamefont {S.}~\bibnamefont {Burov}}, \bibinfo
  {author} {\bibfnamefont {E.}~\bibnamefont {Barkai}}, \bibinfo {author}
  {\bibfnamefont {C.}~\bibnamefont {Selhuber-Unkel}}, \bibinfo {author}
  {\bibfnamefont {K.}~\bibnamefont {Berg-S\o{}rensen}}, \bibinfo {author}
  {\bibfnamefont {L.}~\bibnamefont {Oddershede}}, \ and\ \bibinfo {author}
  {\bibfnamefont {R.}~\bibnamefont {Metzler}},\ }\href {\doibase
  10.1103/PhysRevLett.106.048103} {\bibfield  {journal} {\bibinfo  {journal}
  {Phys. Rev. Lett.}\ }\textbf {\bibinfo {volume} {106}},\ \bibinfo {pages}
  {048103} (\bibinfo {year} {2011})}\BibitemShut {NoStop}%
\bibitem [{\citenamefont {Weigel}\ \emph {et~al.}(2011)\citenamefont {Weigel},
  \citenamefont {Simon}, \citenamefont {Tamkun},\ and\ \citenamefont
  {Krapf}}]{Weigel2011}%
  \BibitemOpen
  \bibfield  {author} {\bibinfo {author} {\bibfnamefont {A.~V.}\ \bibnamefont
  {Weigel}}, \bibinfo {author} {\bibfnamefont {B.}~\bibnamefont {Simon}},
  \bibinfo {author} {\bibfnamefont {M.~M.}\ \bibnamefont {Tamkun}}, \ and\
  \bibinfo {author} {\bibfnamefont {D.}~\bibnamefont {Krapf}},\ }\href
  {\doibase 10.1073/pnas.1016325108} {\bibfield  {journal} {\bibinfo  {journal}
  {Proc. Natl. Acad. Sci. USA}\ }\textbf {\bibinfo {volume} {108}},\ \bibinfo
  {pages} {6438} (\bibinfo {year} {2011})}\BibitemShut {NoStop}%
\bibitem [{\citenamefont {Kusumi}\ \emph {et~al.}(2012)\citenamefont {Kusumi},
  \citenamefont {Fujiwara}, \citenamefont {Chadda}, \citenamefont {Xie},
  \citenamefont {Tsunoyama}, \citenamefont {Kalay}, \citenamefont {Kasai},\
  and\ \citenamefont {Suzuki}}]{Kusumi2012}%
  \BibitemOpen
  \bibfield  {author} {\bibinfo {author} {\bibfnamefont {A.}~\bibnamefont
  {Kusumi}}, \bibinfo {author} {\bibfnamefont {T.~K.}\ \bibnamefont
  {Fujiwara}}, \bibinfo {author} {\bibfnamefont {R.}~\bibnamefont {Chadda}},
  \bibinfo {author} {\bibfnamefont {M.}~\bibnamefont {Xie}}, \bibinfo {author}
  {\bibfnamefont {T.~A.}\ \bibnamefont {Tsunoyama}}, \bibinfo {author}
  {\bibfnamefont {Z.}~\bibnamefont {Kalay}}, \bibinfo {author} {\bibfnamefont
  {R.~S.}\ \bibnamefont {Kasai}}, \ and\ \bibinfo {author} {\bibfnamefont
  {K.~G.}\ \bibnamefont {Suzuki}},\ }\href {\doibase
  10.1146/annurev-cellbio-100809-151736} {\bibfield  {journal} {\bibinfo
  {journal} {Annu. Rev. Cell Dev. Biol.}\ }\textbf {\bibinfo {volume} {28}},\
  \bibinfo {pages} {215} (\bibinfo {year} {2012})}\BibitemShut {NoStop}%
\bibitem [{\citenamefont {Saxton}(1993)}]{Saxton1993}%
  \BibitemOpen
  \bibfield  {author} {\bibinfo {author} {\bibfnamefont {M.~J.}\ \bibnamefont
  {Saxton}},\ }\href {\doibase 10.1016/S0006-3495(93)81548-0} {\bibfield
  {journal} {\bibinfo  {journal} {Biophys. J.}\ }\textbf {\bibinfo {volume}
  {64}},\ \bibinfo {pages} {1766} (\bibinfo {year} {1993})}\BibitemShut
  {NoStop}%
\bibitem [{\citenamefont {Saxton}(1997)}]{Saxton1997}%
  \BibitemOpen
  \bibfield  {author} {\bibinfo {author} {\bibfnamefont {M.~J.}\ \bibnamefont
  {Saxton}},\ }\href {\doibase 10.1016/S0006-3495(97)78820-9} {\bibfield
  {journal} {\bibinfo  {journal} {Biophys. J.}\ }\textbf {\bibinfo {volume}
  {72}},\ \bibinfo {pages} {1744} (\bibinfo {year} {1997})}\BibitemShut
  {NoStop}%
\bibitem [{\citenamefont {Leyvraz}\ \emph {et~al.}(1986)\citenamefont
  {Leyvraz}, \citenamefont {Adler}, \citenamefont {Aharony}, \citenamefont
  {Bunde}, \citenamefont {Coniglio}, \citenamefont {Hong}, \citenamefont
  {Stanley},\ and\ \citenamefont {Stauffer}}]{Leyvraz1986}%
  \BibitemOpen
  \bibfield  {author} {\bibinfo {author} {\bibfnamefont {F.}~\bibnamefont
  {Leyvraz}}, \bibinfo {author} {\bibfnamefont {J.}~\bibnamefont {Adler}},
  \bibinfo {author} {\bibfnamefont {A.}~\bibnamefont {Aharony}}, \bibinfo
  {author} {\bibfnamefont {A.}~\bibnamefont {Bunde}}, \bibinfo {author}
  {\bibfnamefont {A.}~\bibnamefont {Coniglio}}, \bibinfo {author}
  {\bibfnamefont {D.~C.}\ \bibnamefont {Hong}}, \bibinfo {author}
  {\bibfnamefont {H.~E.}\ \bibnamefont {Stanley}}, \ and\ \bibinfo {author}
  {\bibfnamefont {D.}~\bibnamefont {Stauffer}},\ }\href
  {http://stacks.iop.org/0305-4470/19/i=17/a=030} {\bibfield  {journal}
  {\bibinfo  {journal} {J. Phys. A}\ }\textbf {\bibinfo {volume} {19}},\
  \bibinfo {pages} {3683} (\bibinfo {year} {1986})}\BibitemShut {NoStop}%
\bibitem [{\citenamefont {Hottovy}\ \emph {et~al.}(2012)\citenamefont
  {Hottovy}, \citenamefont {Volpe},\ and\ \citenamefont {Wehr}}]{Hottovy12}%
  \BibitemOpen
  \bibfield  {author} {\bibinfo {author} {\bibfnamefont {S.}~\bibnamefont
  {Hottovy}}, \bibinfo {author} {\bibfnamefont {G.}~\bibnamefont {Volpe}}, \
  and\ \bibinfo {author} {\bibfnamefont {J.}~\bibnamefont {Wehr}},\ }\href
  {\doibase 10.1007/s10955-012-0418-9} {\bibfield  {journal} {\bibinfo
  {journal} {J. of Stat. Phys.}\ }\textbf {\bibinfo {volume} {146}},\ \bibinfo
  {pages} {762} (\bibinfo {year} {2012})}\BibitemShut {NoStop}%
\bibitem [{\citenamefont {Cherstvy}\ and\ \citenamefont
  {Metzler}(2013)}]{Cherstvy2013a}%
  \BibitemOpen
  \bibfield  {author} {\bibinfo {author} {\bibfnamefont {A.~G.}\ \bibnamefont
  {Cherstvy}}\ and\ \bibinfo {author} {\bibfnamefont {R.}~\bibnamefont
  {Metzler}},\ }\href {\doibase 10.1039/C3CP53056F} {\bibfield  {journal}
  {\bibinfo  {journal} {Phys. Chem. Chem. Phys.}\ }\textbf {\bibinfo {volume}
  {15}},\ \bibinfo {pages} {20220} (\bibinfo {year} {2013})}\BibitemShut
  {NoStop}%
\bibitem [{\citenamefont {Cherstvy}\ \emph {et~al.}(2013)\citenamefont
  {Cherstvy}, \citenamefont {Chechkin},\ and\ \citenamefont
  {Metzler}}]{Cherstvy2013b}%
  \BibitemOpen
  \bibfield  {author} {\bibinfo {author} {\bibfnamefont {A.~G.}\ \bibnamefont
  {Cherstvy}}, \bibinfo {author} {\bibfnamefont {A.~V.}\ \bibnamefont
  {Chechkin}}, \ and\ \bibinfo {author} {\bibfnamefont {R.}~\bibnamefont
  {Metzler}},\ }\href {http://stacks.iop.org/1367-2630/15/i=8/a=083039}
  {\bibfield  {journal} {\bibinfo  {journal} {New J. Phys.}\ }\textbf {\bibinfo
  {volume} {15}},\ \bibinfo {pages} {083039} (\bibinfo {year}
  {2013})}\BibitemShut {NoStop}%
\bibitem [{\citenamefont {Cherstvy}\ \emph {et~al.}(2014)\citenamefont
  {Cherstvy}, \citenamefont {Chechkin},\ and\ \citenamefont
  {Metzler}}]{Cherstvy2014}%
  \BibitemOpen
  \bibfield  {author} {\bibinfo {author} {\bibfnamefont {A.~G.}\ \bibnamefont
  {Cherstvy}}, \bibinfo {author} {\bibfnamefont {A.~V.}\ \bibnamefont
  {Chechkin}}, \ and\ \bibinfo {author} {\bibfnamefont {R.}~\bibnamefont
  {Metzler}},\ }\href {\doibase 10.1039/C3SM52846D} {\bibfield  {journal}
  {\bibinfo  {journal} {Soft Matter}\ }\textbf {\bibinfo {volume} {10}},\
  \bibinfo {pages} {1591} (\bibinfo {year} {2014})}\BibitemShut {NoStop}%
\bibitem [{\citenamefont {Serge}\ \emph {et~al.}(2008)\citenamefont {Serge},
  \citenamefont {Bertaux}, \citenamefont {Rigneault},\ and\ \citenamefont
  {Marguet}}]{Serge2008}%
  \BibitemOpen
  \bibfield  {author} {\bibinfo {author} {\bibfnamefont {A.}~\bibnamefont
  {Serge}}, \bibinfo {author} {\bibfnamefont {N.}~\bibnamefont {Bertaux}},
  \bibinfo {author} {\bibfnamefont {H.}~\bibnamefont {Rigneault}}, \ and\
  \bibinfo {author} {\bibfnamefont {D.}~\bibnamefont {Marguet}},\ }\href
  {\doibase 10.1038/nmeth.1233} {\bibfield  {journal} {\bibinfo  {journal}
  {Nature Methods}\ }\textbf {\bibinfo {volume} {5}},\ \bibinfo {pages} {687}
  (\bibinfo {year} {2008})}\BibitemShut {NoStop}%
\bibitem [{\citenamefont {English}\ \emph {et~al.}(2011)\citenamefont
  {English}, \citenamefont {Hauryliuk}, \citenamefont {Sanamrad}, \citenamefont
  {Tankov}, \citenamefont {Dekker},\ and\ \citenamefont {Elf}}]{English2011}%
  \BibitemOpen
  \bibfield  {author} {\bibinfo {author} {\bibfnamefont {B.~P.}\ \bibnamefont
  {English}}, \bibinfo {author} {\bibfnamefont {V.}~\bibnamefont {Hauryliuk}},
  \bibinfo {author} {\bibfnamefont {A.}~\bibnamefont {Sanamrad}}, \bibinfo
  {author} {\bibfnamefont {S.}~\bibnamefont {Tankov}}, \bibinfo {author}
  {\bibfnamefont {N.~H.}\ \bibnamefont {Dekker}}, \ and\ \bibinfo {author}
  {\bibfnamefont {J.}~\bibnamefont {Elf}},\ }\href {\doibase
  10.1073/pnas.1102255108} {\bibfield  {journal} {\bibinfo  {journal} {Proc.
  Natl. Acad. Sci. USA}\ }\textbf {\bibinfo {volume} {108}},\ \bibinfo {pages}
  {E365} (\bibinfo {year} {2011})}\BibitemShut {NoStop}%
\bibitem [{\citenamefont {K\"{u}hn}\ \emph {et~al.}(2011)\citenamefont
  {K\"{u}hn}, \citenamefont {Ihalainen}, \citenamefont {Hyv\"{a}luoma},
  \citenamefont {Dross}, \citenamefont {Willman}, \citenamefont {Langowski},
  \citenamefont {Vihinen-Ranta},\ and\ \citenamefont {Timonen}}]{Kuehn2011}%
  \BibitemOpen
  \bibfield  {author} {\bibinfo {author} {\bibfnamefont {T.}~\bibnamefont
  {K\"{u}hn}}, \bibinfo {author} {\bibfnamefont {T.~O.}\ \bibnamefont
  {Ihalainen}}, \bibinfo {author} {\bibfnamefont {J.}~\bibnamefont
  {Hyv\"{a}luoma}}, \bibinfo {author} {\bibfnamefont {N.}~\bibnamefont
  {Dross}}, \bibinfo {author} {\bibfnamefont {S.~F.}\ \bibnamefont {Willman}},
  \bibinfo {author} {\bibfnamefont {J.}~\bibnamefont {Langowski}}, \bibinfo
  {author} {\bibfnamefont {M.}~\bibnamefont {Vihinen-Ranta}}, \ and\ \bibinfo
  {author} {\bibfnamefont {J.}~\bibnamefont {Timonen}},\ }\href {\doibase
  10.1371/journal.pone.0022962} {\bibfield  {journal} {\bibinfo  {journal}
  {PLoS ONE}\ }\textbf {\bibinfo {volume} {6}},\ \bibinfo {pages} {e22962}
  (\bibinfo {year} {2011})}\BibitemShut {NoStop}%
\bibitem [{\citenamefont {Cutler}\ \emph {et~al.}(2013)\citenamefont {Cutler},
  \citenamefont {Malik}, \citenamefont {Liu}, \citenamefont {Byars},
  \citenamefont {Lidke},\ and\ \citenamefont {Lidke}}]{Cutler13}%
  \BibitemOpen
  \bibfield  {author} {\bibinfo {author} {\bibfnamefont {P.~J.}\ \bibnamefont
  {Cutler}}, \bibinfo {author} {\bibfnamefont {M.~D.}\ \bibnamefont {Malik}},
  \bibinfo {author} {\bibfnamefont {S.}~\bibnamefont {Liu}}, \bibinfo {author}
  {\bibfnamefont {J.~M.}\ \bibnamefont {Byars}}, \bibinfo {author}
  {\bibfnamefont {D.~S.}\ \bibnamefont {Lidke}}, \ and\ \bibinfo {author}
  {\bibfnamefont {K.~A.}\ \bibnamefont {Lidke}},\ }\href {\doibase
  10.1371/journal.pone.0064320} {\bibfield  {journal} {\bibinfo  {journal}
  {PLoS ONE}\ }\textbf {\bibinfo {volume} {8}},\ \bibinfo {pages} {e64320}
  (\bibinfo {year} {2013})}\BibitemShut {NoStop}%
\bibitem [{\citenamefont {Giannone}\ \emph {et~al.}(2013)\citenamefont
  {Giannone}, \citenamefont {Hosy}, \citenamefont {Sibarita}, \citenamefont
  {Choquet},\ and\ \citenamefont {Cognet}}]{Giannone13}%
  \BibitemOpen
  \bibfield  {author} {\bibinfo {author} {\bibfnamefont {G.}~\bibnamefont
  {Giannone}}, \bibinfo {author} {\bibfnamefont {E.}~\bibnamefont {Hosy}},
  \bibinfo {author} {\bibfnamefont {J.-B.}\ \bibnamefont {Sibarita}}, \bibinfo
  {author} {\bibfnamefont {D.}~\bibnamefont {Choquet}}, \ and\ \bibinfo
  {author} {\bibfnamefont {L.}~\bibnamefont {Cognet}},\ }in\ \href {\doibase
  10.1007/978-1-62703-137-0_7} {\emph {\bibinfo {booktitle} {Nanoimaging}}},\
  \bibinfo {series} {Methods in Molecular Biology}, Vol.\ \bibinfo {volume}
  {950},\ \bibinfo {editor} {edited by\ \bibinfo {editor} {\bibfnamefont
  {A.~A.}\ \bibnamefont {Sousa}}\ and\ \bibinfo {editor} {\bibfnamefont
  {M.~J.}\ \bibnamefont {Kruhlak}}}\ (\bibinfo  {publisher} {Humana Press},\
  \bibinfo {address} {New York},\ \bibinfo {year} {2013})\ pp.\ \bibinfo
  {pages} {95--110}\BibitemShut {NoStop}%
\bibitem [{\citenamefont {Masson}\ \emph {et~al.}(2014)\citenamefont {Masson},
  \citenamefont {Dionne}, \citenamefont {Salvatico}, \citenamefont {Renner},
  \citenamefont {Specht}, \citenamefont {Triller},\ and\ \citenamefont
  {Dahan}}]{Masson2014}%
  \BibitemOpen
  \bibfield  {author} {\bibinfo {author} {\bibfnamefont {J.~B.}\ \bibnamefont
  {Masson}}, \bibinfo {author} {\bibfnamefont {P.}~\bibnamefont {Dionne}},
  \bibinfo {author} {\bibfnamefont {C.}~\bibnamefont {Salvatico}}, \bibinfo
  {author} {\bibfnamefont {M.}~\bibnamefont {Renner}}, \bibinfo {author}
  {\bibfnamefont {C.}~\bibnamefont {Specht}}, \bibinfo {author} {\bibfnamefont
  {A.}~\bibnamefont {Triller}}, \ and\ \bibinfo {author} {\bibfnamefont
  {M.}~\bibnamefont {Dahan}},\ }\href {\doibase 10.1016/j.bpj.2013.10.027}
  {\bibfield  {journal} {\bibinfo  {journal} {Biophys. J.}\ }\textbf {\bibinfo
  {volume} {106}},\ \bibinfo {pages} {74} (\bibinfo {year} {2014})}\BibitemShut
  {NoStop}%
\bibitem [{\citenamefont {Khoury}\ \emph {et~al.}(2011)\citenamefont {Khoury},
  \citenamefont {Lacasta}, \citenamefont {Sancho},\ and\ \citenamefont
  {Lindenberg}}]{Khoury2011}%
  \BibitemOpen
  \bibfield  {author} {\bibinfo {author} {\bibfnamefont {M.}~\bibnamefont
  {Khoury}}, \bibinfo {author} {\bibfnamefont {A.~M.}\ \bibnamefont {Lacasta}},
  \bibinfo {author} {\bibfnamefont {J.~M.}\ \bibnamefont {Sancho}}, \ and\
  \bibinfo {author} {\bibfnamefont {K.}~\bibnamefont {Lindenberg}},\ }\href
  {\doibase 10.1103/PhysRevLett.106.090602} {\bibfield  {journal} {\bibinfo
  {journal} {Phys. Rev. Lett.}\ }\textbf {\bibinfo {volume} {106}},\ \bibinfo
  {pages} {090602} (\bibinfo {year} {2011})}\BibitemShut {NoStop}%
\bibitem [{\citenamefont {Yuste}\ and\ \citenamefont
  {Lindenberg}(2007)}]{Yuste2007}%
  \BibitemOpen
  \bibfield  {author} {\bibinfo {author} {\bibfnamefont {S.~B.}\ \bibnamefont
  {Yuste}}\ and\ \bibinfo {author} {\bibfnamefont {K.}~\bibnamefont
  {Lindenberg}},\ }\href {\doibase 10.1103/PhysRevE.76.051114} {\bibfield
  {journal} {\bibinfo  {journal} {Phys. Rev. E}\ }\textbf {\bibinfo {volume}
  {76}},\ \bibinfo {pages} {051114} (\bibinfo {year} {2007})}\BibitemShut
  {NoStop}%
\bibitem [{\citenamefont {Condamin}\ \emph {et~al.}(2008)\citenamefont
  {Condamin}, \citenamefont {Tejedor}, \citenamefont {Voituriez}, \citenamefont
  {Bénichou},\ and\ \citenamefont {Klafter}}]{Condamin2008}%
  \BibitemOpen
  \bibfield  {author} {\bibinfo {author} {\bibfnamefont {S.}~\bibnamefont
  {Condamin}}, \bibinfo {author} {\bibfnamefont {V.}~\bibnamefont {Tejedor}},
  \bibinfo {author} {\bibfnamefont {R.}~\bibnamefont {Voituriez}}, \bibinfo
  {author} {\bibfnamefont {O.}~\bibnamefont {Bénichou}}, \ and\ \bibinfo
  {author} {\bibfnamefont {J.}~\bibnamefont {Klafter}},\ }\href {\doibase
  10.1073/pnas.0712158105} {\bibfield  {journal} {\bibinfo  {journal} {Proc.
  Natl. Acad. Sci. USA}\ }\textbf {\bibinfo {volume} {105}},\ \bibinfo {pages}
  {5675} (\bibinfo {year} {2008})}\BibitemShut {NoStop}%
\bibitem [{\citenamefont {Bakker}\ \emph {et~al.}(2012)\citenamefont {Bakker},
  \citenamefont {Eich}, \citenamefont {Torreno-Pina}, \citenamefont
  {Diez-Ahedo}, \citenamefont {Perez-Samper}, \citenamefont {van Zanten},
  \citenamefont {Figdor}, \citenamefont {Cambi},\ and\ \citenamefont
  {Garcia-Parajo}}]{Bakker2012}%
  \BibitemOpen
  \bibfield  {author} {\bibinfo {author} {\bibfnamefont {G.~J.}\ \bibnamefont
  {Bakker}}, \bibinfo {author} {\bibfnamefont {C.}~\bibnamefont {Eich}},
  \bibinfo {author} {\bibfnamefont {J.~A.}\ \bibnamefont {Torreno-Pina}},
  \bibinfo {author} {\bibfnamefont {R.}~\bibnamefont {Diez-Ahedo}}, \bibinfo
  {author} {\bibfnamefont {G.}~\bibnamefont {Perez-Samper}}, \bibinfo {author}
  {\bibfnamefont {T.~S.}\ \bibnamefont {van Zanten}}, \bibinfo {author}
  {\bibfnamefont {C.~G.}\ \bibnamefont {Figdor}}, \bibinfo {author}
  {\bibfnamefont {A.}~\bibnamefont {Cambi}}, \ and\ \bibinfo {author}
  {\bibfnamefont {M.~F.}\ \bibnamefont {Garcia-Parajo}},\ }\href {\doibase
  10.1073/pnas.1116425109} {\bibfield  {journal} {\bibinfo  {journal} {Proc.
  Natl. Acad. Sci. USA}\ }\textbf {\bibinfo {volume} {109}},\ \bibinfo {pages}
  {4869} (\bibinfo {year} {2012})}\BibitemShut {NoStop}%
\bibitem [{\citenamefont {Rossier}\ \emph {et~al.}(2012)\citenamefont
  {Rossier}, \citenamefont {Octeau}, \citenamefont {Sibarita}, \citenamefont
  {Leduc}, \citenamefont {Tessier}, \citenamefont {Nair}, \citenamefont
  {Gatterdam}, \citenamefont {Destaing}, \citenamefont {Albiges-Rizo},
  \citenamefont {Tampe}, \citenamefont {Cognet}, \citenamefont {Choquet},
  \citenamefont {Lounis},\ and\ \citenamefont {Giannone}}]{Rossier2012}%
  \BibitemOpen
  \bibfield  {author} {\bibinfo {author} {\bibfnamefont {O.}~\bibnamefont
  {Rossier}}, \bibinfo {author} {\bibfnamefont {V.}~\bibnamefont {Octeau}},
  \bibinfo {author} {\bibfnamefont {J.-B.}\ \bibnamefont {Sibarita}}, \bibinfo
  {author} {\bibfnamefont {C.}~\bibnamefont {Leduc}}, \bibinfo {author}
  {\bibfnamefont {B.}~\bibnamefont {Tessier}}, \bibinfo {author} {\bibfnamefont
  {D.}~\bibnamefont {Nair}}, \bibinfo {author} {\bibfnamefont {V.}~\bibnamefont
  {Gatterdam}}, \bibinfo {author} {\bibfnamefont {O.}~\bibnamefont {Destaing}},
  \bibinfo {author} {\bibfnamefont {C.}~\bibnamefont {Albiges-Rizo}}, \bibinfo
  {author} {\bibfnamefont {R.}~\bibnamefont {Tampe}}, \bibinfo {author}
  {\bibfnamefont {L.}~\bibnamefont {Cognet}}, \bibinfo {author} {\bibfnamefont
  {D.}~\bibnamefont {Choquet}}, \bibinfo {author} {\bibfnamefont
  {B.}~\bibnamefont {Lounis}}, \ and\ \bibinfo {author} {\bibfnamefont
  {G.}~\bibnamefont {Giannone}},\ }\href {\doibase 10.1038/ncb2620} {\bibfield
  {journal} {\bibinfo  {journal} {Nat. Cell Biol.}\ }\textbf {\bibinfo {volume}
  {14}},\ \bibinfo {pages} {1231} (\bibinfo {year} {2012})}\BibitemShut
  {NoStop}%
\bibitem [{\citenamefont {Sezgin}\ \emph {et~al.}(2012)\citenamefont {Sezgin},
  \citenamefont {Levental}, \citenamefont {Grzybek}, \citenamefont
  {Schwarzmann}, \citenamefont {Mueller}, \citenamefont {Honigmann},
  \citenamefont {Belov}, \citenamefont {Eggeling}, \citenamefont {Coskun},
  \citenamefont {Simons},\ and\ \citenamefont {Schwille}}]{Sezgin2012}%
  \BibitemOpen
  \bibfield  {author} {\bibinfo {author} {\bibfnamefont {E.}~\bibnamefont
  {Sezgin}}, \bibinfo {author} {\bibfnamefont {I.}~\bibnamefont {Levental}},
  \bibinfo {author} {\bibfnamefont {M.}~\bibnamefont {Grzybek}}, \bibinfo
  {author} {\bibfnamefont {G.}~\bibnamefont {Schwarzmann}}, \bibinfo {author}
  {\bibfnamefont {V.}~\bibnamefont {Mueller}}, \bibinfo {author} {\bibfnamefont
  {A.}~\bibnamefont {Honigmann}}, \bibinfo {author} {\bibfnamefont {V.~N.}\
  \bibnamefont {Belov}}, \bibinfo {author} {\bibfnamefont {C.}~\bibnamefont
  {Eggeling}}, \bibinfo {author} {\bibfnamefont {{\"{U}}.}~\bibnamefont
  {Coskun}}, \bibinfo {author} {\bibfnamefont {K.}~\bibnamefont {Simons}}, \
  and\ \bibinfo {author} {\bibfnamefont {P.}~\bibnamefont {Schwille}},\ }\href
  {\doibase http://dx.doi.org/10.1016/j.bbamem.2012.03.007} {\bibfield
  {journal} {\bibinfo  {journal} {Biochim. Biophys. Acta}\ }\textbf {\bibinfo
  {volume} {1818}},\ \bibinfo {pages} {1777 } (\bibinfo {year}
  {2012})}\BibitemShut {NoStop}%
\bibitem [{\citenamefont {Shlesinger}\ \emph {et~al.}(1987)\citenamefont
  {Shlesinger}, \citenamefont {West},\ and\ \citenamefont
  {Klafter}}]{Shlesinger1987}%
  \BibitemOpen
  \bibfield  {author} {\bibinfo {author} {\bibfnamefont {M.~F.}\ \bibnamefont
  {Shlesinger}}, \bibinfo {author} {\bibfnamefont {B.~J.}\ \bibnamefont
  {West}}, \ and\ \bibinfo {author} {\bibfnamefont {J.}~\bibnamefont
  {Klafter}},\ }\href {\doibase 10.1103/PhysRevLett.58.1100} {\bibfield
  {journal} {\bibinfo  {journal} {Phys. Rev. Lett.}\ }\textbf {\bibinfo
  {volume} {58}},\ \bibinfo {pages} {1100} (\bibinfo {year}
  {1987})}\BibitemShut {NoStop}%
\bibitem [{\citenamefont {Klafter}\ \emph {et~al.}(1987)\citenamefont
  {Klafter}, \citenamefont {Blumen},\ and\ \citenamefont
  {Shlesinger}}]{Klafter1987}%
  \BibitemOpen
  \bibfield  {author} {\bibinfo {author} {\bibfnamefont {J.}~\bibnamefont
  {Klafter}}, \bibinfo {author} {\bibfnamefont {A.}~\bibnamefont {Blumen}}, \
  and\ \bibinfo {author} {\bibfnamefont {M.~F.}\ \bibnamefont {Shlesinger}},\
  }\href {\doibase 10.1103/PhysRevA.35.3081} {\bibfield  {journal} {\bibinfo
  {journal} {Phys. Rev. A}\ }\textbf {\bibinfo {volume} {35}},\ \bibinfo
  {pages} {3081} (\bibinfo {year} {1987})}\BibitemShut {NoStop}%
\bibitem [{Note1()}]{Note1}%
  \BibitemOpen
  \bibinfo {note} {When writing probability densities and probabilities, we do
  not distinguish between arguments representing values of random variables and
  other parameters. However, we do write the former before the
  latter.}\BibitemShut {Stop}%
\bibitem [{\citenamefont {Feller}(1971)}]{feller-vol-2}%
  \BibitemOpen
  \bibfield  {author} {\bibinfo {author} {\bibfnamefont {W.}~\bibnamefont
  {Feller}},\ }\href@noop {} {\emph {\bibinfo {title} {An introduction to
  probability theory and its applications. {V}ol. {II}.}}},\ Second edition\
  (\bibinfo  {publisher} {John Wiley \& Sons Inc.},\ \bibinfo {address} {New
  York},\ \bibinfo {year} {1971})\BibitemShut {NoStop}%
\bibitem [{\citenamefont {Machta}(1985)}]{Machta85}%
  \BibitemOpen
  \bibfield  {author} {\bibinfo {author} {\bibfnamefont {J.}~\bibnamefont
  {Machta}},\ }\href {\doibase 10.1088/0305-4470/18/9/008} {\bibfield
  {journal} {\bibinfo  {journal} {J. Phys. A}\ }\textbf {\bibinfo {volume}
  {18}},\ \bibinfo {pages} {L531} (\bibinfo {year} {1985})}\BibitemShut
  {NoStop}%
\bibitem [{\citenamefont {{J. P. Bouchaud}}(1992)}]{Bouchaud1992}%
  \BibitemOpen
  \bibfield  {author} {\bibinfo {author} {\bibnamefont {{J. P. Bouchaud}}},\
  }\href {\doibase 10.1051/jp1:1992238} {\bibfield  {journal} {\bibinfo
  {journal} {J. Phys. I France}\ }\textbf {\bibinfo {volume} {2}},\ \bibinfo
  {pages} {1705} (\bibinfo {year} {1992})}\BibitemShut {NoStop}%
\bibitem [{\citenamefont {Durrett}(1996)}]{Durrett96}%
  \BibitemOpen
  \bibfield  {author} {\bibinfo {author} {\bibfnamefont {R.}~\bibnamefont
  {Durrett}},\ }\href@noop {} {\emph {\bibinfo {title} {Stochastic Calculus: A
  Practical Introduction (Probability and Stochastics Series)}}}\ (\bibinfo
  {publisher} {CRC Press},\ \bibinfo {address} {Boca Raton},\ \bibinfo {year}
  {1996})\BibitemShut {NoStop}%
\bibitem [{\citenamefont {Mandelbrot}\ and\ \citenamefont
  {Ness}(1968)}]{Mandelbrot1968}%
  \BibitemOpen
  \bibfield  {author} {\bibinfo {author} {\bibfnamefont {B.~B.}\ \bibnamefont
  {Mandelbrot}}\ and\ \bibinfo {author} {\bibfnamefont {J.~W.~V.}\ \bibnamefont
  {Ness}},\ }\href {\doibase 10.1137/1010093} {\bibfield  {journal} {\bibinfo
  {journal} {SIAM Rev.}\ }\textbf {\bibinfo {volume} {10}},\ \bibinfo {pages}
  {422} (\bibinfo {year} {1968})}\BibitemShut {NoStop}%
\bibitem [{\citenamefont {Barkai}\ and\ \citenamefont
  {Sokolov}(2007)}]{Barkai2007}%
  \BibitemOpen
  \bibfield  {author} {\bibinfo {author} {\bibfnamefont {E.}~\bibnamefont
  {Barkai}}\ and\ \bibinfo {author} {\bibfnamefont {I.~M.}\ \bibnamefont
  {Sokolov}},\ }\href {\doibase 10.1088/1742-5468/2007/08/P08001} {\bibfield
  {journal} {\bibinfo  {journal} {J. Stat. Mech.}\ }\textbf {\bibinfo {volume}
  {2007}},\ \bibinfo {pages} {P08001} (\bibinfo {year} {2007})}\BibitemShut
  {NoStop}%
\bibitem [{\citenamefont {Ben~Arous}\ and\ \citenamefont
  {\u{C}ern\'y}(2007)}]{benarous2007}%
  \BibitemOpen
  \bibfield  {author} {\bibinfo {author} {\bibfnamefont {G.}~\bibnamefont
  {Ben~Arous}}\ and\ \bibinfo {author} {\bibfnamefont {J.}~\bibnamefont
  {\u{C}ern\'y}},\ }\href {\doibase 10.1214/009117907000000024} {\bibfield
  {journal} {\bibinfo  {journal} {Ann. Probab.}\ }\textbf {\bibinfo {volume}
  {35}},\ \bibinfo {pages} {2356} (\bibinfo {year} {2007})}\BibitemShut
  {NoStop}%
\end{thebibliography}%
\end{document}